\documentclass[conference]{IEEEtran}
\usepackage{fancyhdr}
\IEEEoverridecommandlockouts
\usepackage[noadjust]{cite}
\usepackage{amsmath,amssymb,amsfonts}
\usepackage{algorithm } 
\usepackage{graphicx}
\usepackage{textcomp}
\usepackage{xcolor}
\usepackage{hyperref}
\usepackage{cleveref}
\usepackage[noend]{algpseudocode}
\usepackage{subfig}

\usepackage{inconsolata}
\usepackage{listings}
\usepackage{moresize}

\usepackage{multicol}
\usepackage{url}

\usepackage{caption}
\def\BibTeX{{\rm B\kern-.05em{\sc i\kern-.025em b}\kern-.08em
    T\kern-.1667em\lower.7ex\hbox{E}\kern-.125emX}}

\newcommand{\modelname}[0]{SMaT}
\newcommand{\newparagraph}[1]{\noindent\textbf{#1\hspace{1em}}}

\begin{document}

\title{High Performance Unstructured SpMM Computation Using Tensor Cores\\
}

\author{
\IEEEauthorblockN{Patrik Okanovic}
\IEEEauthorblockA{ETH Zurich \\
Department of Computer Science\\
Zurich, Switzerland \\
patrik.okanovic@inf.ethz.ch}\\   
\IEEEauthorblockN{Maciej Besta}
\IEEEauthorblockA{ETH Zurich \\
Department of Computer Science\\
Zurich, Switzerland \\
maciej.besta@inf.ethz.ch}
\and
\IEEEauthorblockN{Grzegorz Kwasniewski}
\IEEEauthorblockA{ETH Zurich \\
Department of Computer Science\\
Zurich, Switzerland \\
gkwasnie@inf.ethz.ch}\\  
\IEEEauthorblockN{Flavio Vella}
\IEEEauthorblockA{University of Trento \\
Trento, Italy \\
flavio.vella@unitn.it}
\and
\IEEEauthorblockN{Paolo Sylos Labini}
\IEEEauthorblockA{Free University of Bozen-Bolzano \\
Faculty of Engineering \\
Bolzano, Italy \\
Paolo.SylosLabini@student.unibz.it}\\                 
\IEEEauthorblockN{Torsten Hoefler}
\IEEEauthorblockA{ETH Zurich \\
Department of Computer Science\\
Zurich, Switzerland \\
htor@inf.ethz.ch}
}

\maketitle
\thispagestyle{fancy}
\lhead{}
\rhead{}
\chead{}
\lfoot{\footnotesize{
SC24, November 17-22, 2024, Atlanta, Georgia, USA
\newline 979-8-3503-5291-7/24/\$31.00 \copyright 2024 IEEE}}
\rfoot{}
\cfoot{}
\renewcommand{\headrulewidth}{0pt}
\renewcommand{\footrulewidth}{0pt}

\begin{abstract}
High-performance sparse matrix--matrix (SpMM) multiplication is paramount for science and industry, as the ever-increasing sizes of data prohibit using dense data structures. Yet, existing hardware, such as Tensor Cores (TC), is ill-suited for SpMM, as it imposes strict constraints on data structures that cannot be met by unstructured sparsity found in many applications. To address this, we introduce (S)parse (Ma)trix Matrix (T)ensor Core-accelerated (\modelname): a novel SpMM library that utilizes TCs for unstructured sparse matrices. Our block-sparse library leverages the low-level CUDA MMA (matrix-matrix-accumulate) API, maximizing the performance offered by modern GPUs. Algorithmic optimizations such as sparse matrix permutation, further improve performance by minimizing the number of non-zero blocks. The evaluation on NVIDIA A100 shows that \modelname{} outperforms SotA libraries (DASP, cuSPARSE, and Magicube) by up to 125x (on average 2.6x). \modelname{} can be used to accelerate many workloads in scientific computing, large model training, inference, and others.

\end{abstract}


\section{Introduction}

\begin{figure*}
\centerline{\includegraphics[width=2.1\columnwidth]{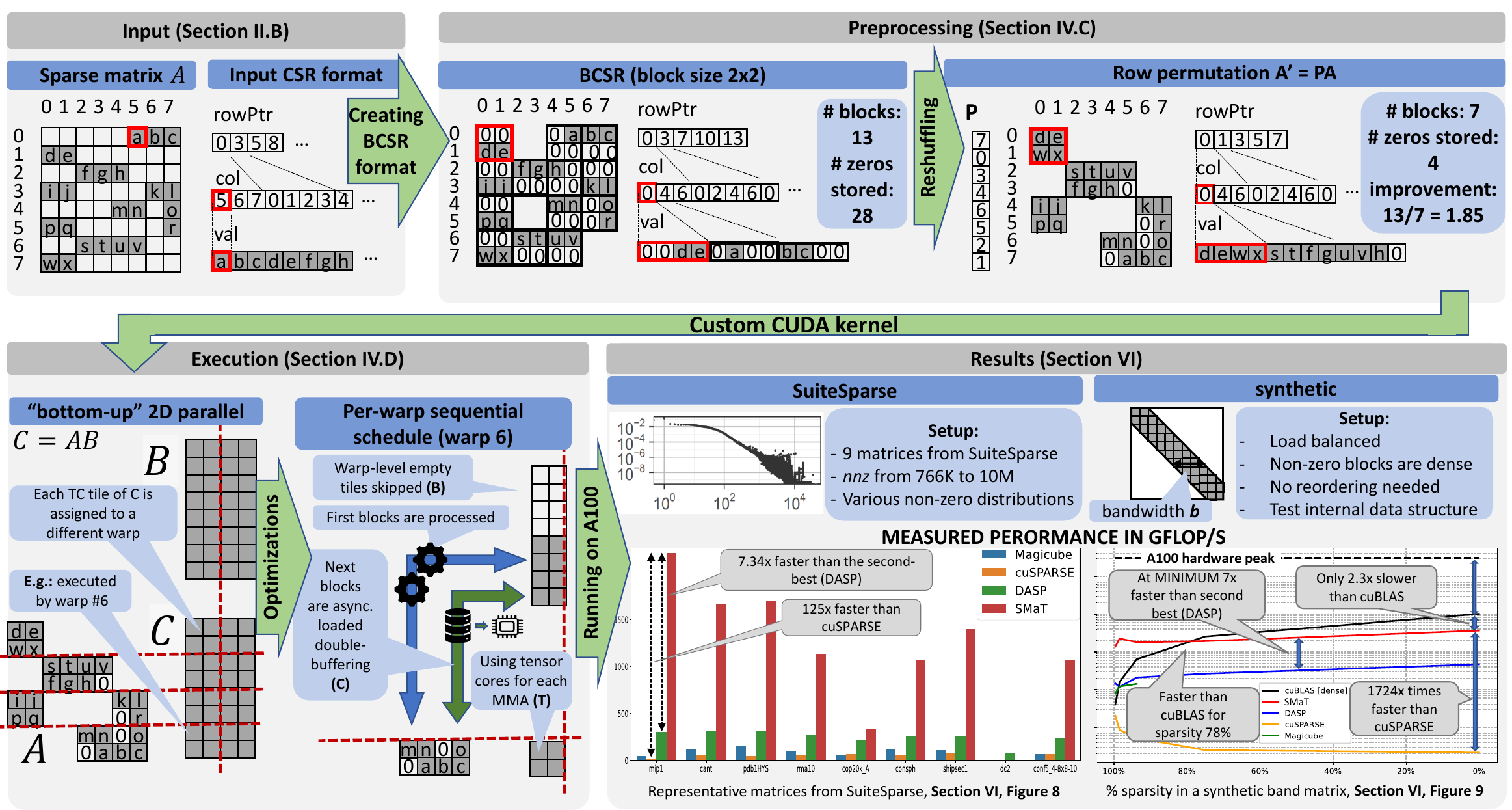}}
\caption{A bird's-eye view of the entire \modelname{}'s pipeline. \modelname{} performs SpMM on an input matrix in the CSR format stored in
any precision supported by Tensor Cores. Then, it preprocesses the matrix to maximize the block density, minimize the total number of blocks,
and maximize the load balance between rows. The preprocessing is done only once and the matrix is internally stored in the BCSR (Blocked-CSR) format.
When the SpMM kernel is launched, an optimized CUDA kernel uses block-level bottom-up 2D parallelism to maximize the utilization of GPU hardware 
resources. The results - both on the SuiteSparse and on synthetic matrices confirm the universality of \modelname{}: it significantly outperforms
remaining solutions in almost every test case: for very sparse and relatively dense matrices, for highly unstructured and for very regular matrices.}
\label{fig:overview}
\end{figure*}

The performance of dense matrix multiplication has steadily improved in recent years, with new architectures and libraries feeding the growing computational needs of deep learning. \emph{MMA} units such as the Tensor Processing Unit~(TPU)~\cite{jouppi2017datacenter} and the Tensor Core~(TC)~\cite{abdelkhalik2022demystifying} are designed to efficiently handle big volumes of \emph{multiply-accumulate\/} operations, a fundamental workload in scientific computing.
However, the performance of multiplying a \textit{sparse} matrix with a dense one (SpMM),
which is not only crucial for important High-Performance workloads~\cite{bai2000templates} or graph and data 
analytics~\cite{gilbert2008unified, besta2017slimsell, besta2017push, solomonik2017scaling, besta2019graph} 
but also constitutes a growing part of many modern workloads---most notably,
Graph Neural Networks (GNNs)~\cite{wu2020comprehensive, survGNN-acc-surv, besta2023parallel, bazinska2023cached, besta2023high} and general Sparse Deep Learning~\cite{hoefler2021sparsity, sparseConvNet}, is far from reaching hardware peak performance.

It is still unclear how to leverage the hardware (HW) and software (SW) machinery developed for dense matrix computations 
in applications operating on sparse matrices. Thus, the computational power of dense matrix units, which are common in most high-performance hardware configurations, still remains untapped. 
Thanks to the efficient use of memory hierarchies and matrix units, explicitly storing zeros and using dense representation can be faster than computations on sparse data structures, even on very sparse matrices.
For example, on the NVIDIA HW/SW stack, the sparsity threshold for the supremacy of dense multiplication (cuBLAS) over sparse SpMM (cuSPARSE) lays as high as $99.9\%~$\cite{sparsert}, depending on the size of the matrix and its sparsity pattern. Multiplying sparse matrices as if they were dense, however, is inherently inefficient due to \textit{padding}---explicitly stored zeros. As the size and sparsity of the matrices grow, padding increases the storage requirements and data movement costs compared to sparse storage, relegating these powerful routines to small matrices. Most importantly, padding reduces the utilization of matrix units, wastefully processing null elements.

A natural approach to reduce padding and increase the utilization of matrix units, and thus to make the SpMM based on dense matrix units faster and more memory-efficient, is \emph{blocking}---tiling a sparse matrix into blocks, and only storing and processing the nonzero ones~\cite{FilipponeReview}. 
Blocking enables partitioning a sparse matrix into a collection of dense matrices---the nonzero blocks---which can then be multiplied using dense units.
However, two main challenges make reaching peak hardware performance of blocked-sparse matrix multiplication still an open problem: a) \textbf{block density}: 
finding the optimal blocking that minimizes the amount of padding - the number of explicit zeros per each dense block that increase memory footprint and account for wasted arithmetic operations;
and b) \textbf{hardware-aware implementation}:
efficient streaming of consecutive blocks to fully utilize memory pipelines, hiding the load latency and saturating all TC units.§

While the problem of efficient utilization of MMA units for sparse matrices was addressed in previous research, existing work mostly focus on narrow use-cases. 
Magicube~\cite{Magicube} is an SpMM library specifically designed for deep learning, requiring structured sparsity and supports only low-precision integers. 
DASP~\cite{lu2023dasp} supports only SpMV operation, which can be viewed as a special case of SpMM, with the dense matrix containing only a single column. 
NVIDIA's general-purpose cuSPARSE library~\cite{naumov2010cusparse} relies on a CSR format, but experiments show that it underperforms in many 
scenarios~\cite{chen2021efficient, Magicube, lu2023dasp}. cuSPARSELt and VENOM~\cite{castro2023venom} work only on fixed sparsity patterns.

In this work, we introduce \emph{\modelname} --- (S)parse (Ma)trix Matrix (T)ensor Core-accelerated library. It is a \emph{general-purpose} SpMM library that works on unstructured sparse matrices in CSR format - arguably the most prevalent format for this purpose. It works with all data types supported by the MMA hardware units. The library first does the {preprocessing permutation} of the sparse matrix to \textit{minimize the number of dense blocks}. Then, our highly-optimized low-level CUDA implementation \textit{efficiently streams thread-level blocks}, overlapping computation with data movement and saturating all hardware TC units using bottom-up~\cite{cosma} 2D parallelism.

\modelname{}, while being a general-purpose library, significantly outperforms both vendor-optimized cuSPARSE, as well as use-case-specific SotA solutions like Magicube and DASP. We compare against the DASP SpMV library by treating SpMM as a batched SpMV and we show that \modelname\ outperforms DASP for batch size as small as $4$. Our empirical performance model 
(Section~\ref{sec:performance_model}) outlines our driving design choices: hardware-aware, data-movement-centric
code design can bring more performance to contemporary GPU architectures than even sophisticated preprocessing algorithms
aimed to reduce the number of blocks. 

Experiments on synthetic matrices show up to 2,445 times improvement over cuSPARSE,
with at least a minimum of 5.3x improvement compared to the second-best library (Section~\ref{sec:results_synthetic}).
Furthermore, we compared the performance of \modelname{} against cuBLAS: a vendor-optimized GEMM library for \emph{dense} matrices,
and show that contrary to previous findings~\cite{sparsert} we outperform cuBLAS for sparsity regimes as low as 78\%.
Evaluated on the real-world matrices from cuSPARSE, we measure up to $8.6\times$ performance improvement (on average $4.8\times$).
The high-level design together with representative results are presented in Figure~\ref{fig:overview}.

To summarize, we provide the following contributions:

\begin{itemize}
    \item \modelname: an end-to-end solution for general-purpose SpMM that supports unstructured sparsity and all data types supported by the TC units,
    \item A sparse matrix preprocessing permutation scheme that decreases the number of dense blocks by up to 2.5x,
    \item A high-performance implementation of the blocked-CSR (BCSR) SpMM on Tensor Cores using the low-level CUDA MMA API,
        \item An empirical performance model for SpMV in BCSR format that quantifies both the impact of preprocessing and the implementation optimizations,
    \item An evaluation demonstrating speedups of up to 125x over cuSPARSE and up to 7.3x over second-best among tested state-of-the-art SpMM routines: cuSPARSE, Magicube, and DASP.
\end{itemize}

\section{Background}

Throughout the paper, we consider a matrix-matrix multiplication $C= AB$ with $C \in \mathbb{F}^{M \times N}$, 
$A \in \mathbb{F}^{M \times K}$, and $B \in \mathbb{F}^{K \times N}$, for some ring $\mathbb{F}$. Furthermore, we assume
matrix $A$ to be \emph{sparse}: denoting the number of non-zero elements in $A$ as \emph{nnz}, 
we have $1 - nnz/(M\cdot K) = \Omega(1)$. Matrix $B$, of size $K \times N$, is dense. The SpMV operator (sparse Matrix-Vector multiplication) can be seen as a special case with $N=1$. 

We first introduce fundamental concepts.
\subsection{Hardware execution model}

\subsubsection{Execution model}

The execution model leverages a hierarchy that includes threads, warps, and thread blocks. A thread is the smallest unit of execution. A warp is group of threads that are executed simultaneously by the GPU. The size of a warp is specific to the GPU architecture, with 32 being a common size in NVIDIA GPUs. Warps execute independently but can share data with other warps in the same thread block using shared memory. They have their own set of private registers and can be synchronized within their block.

\subsubsection{Tensor Cores}

Tensor Cores (TCs) are specialized processing units found in most modern NVIDIA GPUs. These cores are optimized for performing mixed-precision matrix multiply-and-accumulate operations very efficiently, which are fundamental to the training and inference. TCs perform matrix operations where inputs are typically in lower precision formats (like FP16, BF16, or INT8), but accumulate results in a higher precision format (like FP32) to maintain the precision of the computations. A typical operation performed by a TC involves multiplying two small matrices and adding the result to a third matrix, i.e., Fused Multiply-Add (FMA).

When a CUDA kernel that leverages TCs is executed, the warp scheduler on the GPU is responsible for directing warps to utilize these cores efficiently. The warp scheduler must manage the distribution of matrix operations across the available TCs, ensuring that the workload is evenly distributed and that TCs are kept busy to maximize throughput.

\subsubsection{Memory model}
The CUDA memory model provides a hierarchy of memory options for optimizing performance in NVIDIA GPUs. While this hierarchy is common across GPU generations~\cite{aamodt2018general}, we provide the parameters for the A100-SXM4-40GB architecture~\cite{abdelkhalik2022demystifying}, as this is our primary target in the Evaluation (Section~\ref{sec:evaluation}).
\begin{itemize}
    \item Global memory: large shared space accessible by all threads, yet slower compared to other memory types. \textbf{40GB HBM2, bandwidth 1.5TB/s.}
    \item Shared memory: smaller but significantly faster memory shared by threads within a block, making it suitable for data that requires frequent access by cooperating threads. \textbf{Configurable, up to 164KB per SM, 32 banks with bandwidth 64b per clock cycle.}
    \item Registers: smallest and fastest memory units, private to each thread, ideal for storing frequently used variables within a single thread. \textbf{256KB per SM.}
\end{itemize}
In order to utilize TCs, data needs to be strategically moved from global memory to registers. To avoid bank conflicts and 
efficiently overlap computation with data movement, memory alignment and software pipelining play an important role by hiding
data movement latency. Further details on the implementation can be found in \Cref{sec:implementation}.

\subsection{Sparse matrix representation}
\subsubsection{Unstructured sparsity}
Without any information about the topological structure of nonzeros, each individual element has to contain information not only about its value but also about its location,  yielding $\Omega(nnz)$ additional memory overhead.
CSR (Compressed Sparse Row), CSC (Compressed Sparse Column), and CO (Coordinate) are arguably the most common formats for storing unstructured sparsity, with the first being dominant. Despite the memory overhead for storing locations, they tend to have the smallest memory footprint, as no zero values are stored. That comes at the cost of latency (each value's location has to be individually decoded), load balance, and cache utilization.

    
\subsubsection{Structured sparsity}
Structured sparsity has recently gained significant attention due to its applications in deep learning and the introduction of Sparse Tensor Cores (SPTC) in NVIDIA's Ampere architecture. However, SPTCs only support the 2:4 format, which limits achievable sparsity ratios to 50\%. This format requires every consecutive 4 elements to have 2 nonzero values, promising a 2× speedup. Castro et al. extended the algorithmic support for M:N
sparsity~\cite{castro2023venom} --- among every N consecutive values, M are non-zeros. Low-rank decomposition~\cite{yu2017compressing} can be viewed as the lossy compression format, where a large matrix is represented as a product of lower-rank matrices. Its primary advantage
is that the multiplicand matrices are usually dense and naturally fit for dense GEMM-optimized libraries and hardware. 

In general, the performance of structured sparse matrices is higher due to lower position decoding overhead, streamed and vectorizable access patterns, and hardware support. However, in many applications, such as analysis of
real-world graphs~\cite{survGNN-acc-surv}, higher-order patterns can be prohibitively expensive to discover or may be even non-existent~\cite{besta2021graphminesuite}.
\subsubsection{Blocked format}
\label{sec:blocked_format}
While the unstructured format addresses each individual value in the matrix, the \emph{blocked} format addresses the block granularity: submatrices of fixed size $h \times w$~\cite{eberhardt2016optimization}. Each block is assumed to be dense,
that is, all $h \cdot w$ values are stored explicitly, even if some of them are zeros. While each block can, in principle, start at any row and column offset, in this work, we assume a fixed block structure: each block offset
(the row and column index of the top left value of the block) is a multiple of $h$ or $w$, respectively.
Given a matrix $A \in \mathbb{F}^{n \times m}$ with $n$ rows and $m$ columns, and given the rectangular block sizes $h$ and $w$, we identify $A$ with its blocked version $\textbf{A}$ (omitting the dependency on $h$ and $w$). Specifically,
\textbf{A} is an $N \times M$ block matrix, where $N = \lceil{\frac{n}{h}}\rceil$ and
$M = \lceil{\frac{m}{w}}\rceil$. Each of its elements $\textbf{A}_{i,j}$ is a block in 
$\mathbb{F}^{h\times w}$, containing all entries $A_{k,l}$ s.t. 
$\lfloor{\frac{k}{h}} \rfloor = i$ and $\lfloor{\frac{l}{w}}\rfloor = j$.

Once a matrix has been blocked, it can be stored and processed in a blocked storage format, where only nonzero blocks (those with at least one nonzero) are stored explicitly. We refer to zero entries within a nonzero block as \emph{padding}. 

Storing and processing padding elements is one of the two main costs of blocked SpMM, the other being accessing the nonzero blocks themselves. The former grows with the size of the blocks, while the latter grows with their number. 
On the one hand of this trade-off lay purely sparse storage formats, such as the established Compressed Sparse Rows (CSR), which only store nonzero elements but need to access each one separately. On the other hand stand fully dense storage schemes, which store all zeros and nonzeros in one large block. Block storage formats, such as those considered in this paper, carefully balance between these two extremes.


\section{Performance model}
\label{sec:performance_model}
\begin{figure}[htbp]
\hspace*{-0.2cm}\includegraphics[width=1.05\columnwidth]{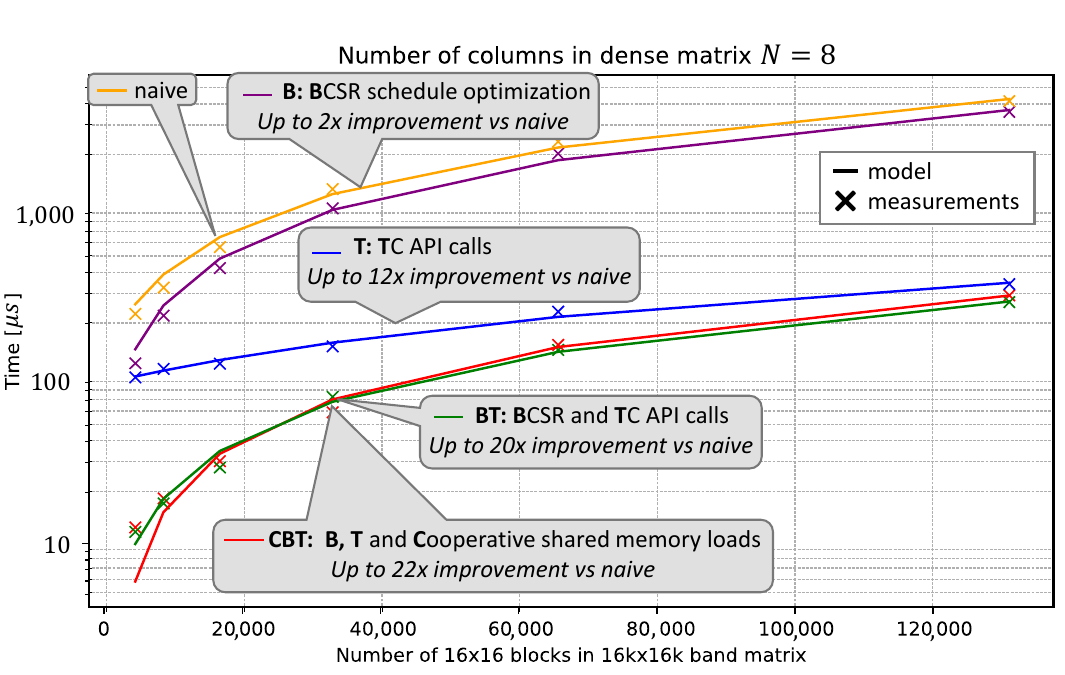}
\caption{Performance measurements vs. model (Equation~\ref{eq:perf}) for various combinations of low-level
optimizations: \textbf{C}: warp-cooperative asynchronous loading from global to shared memory using 
 \texttt{memcpy\_async}; 
 \textbf{B}: using BCSR pointer array to skip empty-block evaluation in the inner loop;
\textbf{T}: using TC MMA API (\texttt{MMA16816})}
\label{fig:perf_model}
\end{figure}

\noindent
 We use a linear performance model 
 \begin{equation}
 \label{eq:perf}
 T_{tot} =  T_{e} \cdot n_e + T_{init},
 \end{equation}
where $T_{tot}$ is the total runtime of the kernel, $T_{e}$ is the execution time of a single compute instruction, 
$n_e$ is the number of elementary computations, and $T_{init}$ accounts for any startup, initialization, 
cache warm-up, and finalization overhead. 
Depending on the implementation, this compute instruction may be, e.g., a single scalar FMA (fused multiply-add), 
a vectorized operation (e.g., using AVX), or a Tensor Core MMA instruction. 

We validate our model and fit the parameters on a $16k \times 16k$ band matrix with varying bandwidth (from 64 to 4096) times a tall-and-skinny dense matrix of size $16k \times 8$. 
We note that such a scenario does not reflect typical real-world sparse matrices, yet we want to isolate any effects
of imperfect load balancing, non-zero distribution, or varying block fill-ins. In this synthetic benchmark, 
our model matches the measurements well (Figure~\ref{fig:perf_model}). 

This model captures two crucial aspects of high-performance SpMM and their equal contribution: low-level implementation $T_e$ and high-level algorithmic optimization $n_e$. 

\subsection{Single instruction time $T_e$}
\label{sec:elem_computation}

\modelname{} fully utilizes TC: thus, depending on the used precision, we treat a single MMA instruction as an elementary computation (e.g., for FP16, we use 16x8x16 MMA).
Optimizing MMM in CUDA requires multiple stages of careful optimizations~\cite{MMM_blogpost}
--- Figure~\ref{fig:perf_model} shows the impact of three representative steps.
In Section~\ref{sec:implementation}, we discuss these optimization in detail.

\subsection{Number of elementary computations $n_e$}
The number of elementary computations $n_e$ is the number of dense blocks in the BCSR format (Section~\ref{sec:blocked_format}). As mentioned in Section~\ref{sec:elem_computation}, the size of the block $h \times w$ depends on the used precision. For a sparse matrix of size $N \times M$ holding $nnz \le N \cdot M$ nonzero values, it can be seen that $n_e$ is bounded by 
\begin{equation}
    \label{eq:num_blocks}
    \frac{nnz}{h \cdot w} \le n_e \le  \min \Big(\frac{N\cdot M}{h \cdot w}, nnz \Big)
\end{equation}

The lower bound is achieved when all blocks are fully packed and there is no zero fill-in. The upper bound represents the case where there is only a single non-zero per entire block, and the remaining $h \cdot w -1$ elements are explicitly stored zeros. Section~\ref{sec:reordering} discusses how \modelname{} reduces the number of blocks $n_e$ by matrix reordering via clustering rows and columns based on the similarity metrics.

\textbf{Observation:} We argue that low-level kernel optimizations can play a more important role than 
even an optimal preprocessing algorithm. While experimenting with different reshuffling algorithms we rarely 
observed a reduction in the total number of blocks greater than 3 times. On the other hand, just using the TC API increases performance by 10 times, with our optimized kernel outperforming a naive implementation 22 times (Figure~\ref{fig:perf_model}). 

\section{\modelname — (S)parse (Ma)trix Matrix (T)ensor Core-accelerated library}
In this section we describe \modelname{}  --- a novel SpMM library designed to utilize TCs for unstructured
sparse matrices. 


\subsection{Overview}
\Cref{fig:overview} displays the overview of \modelname{}.
Initially, the sparse matrix $A$ is read in the CSR format. Internally, the matrix is converted to the 
block format. Afterwards, in the preprocessing phase, the matrix is reshuffled to minimize the number of non-zero blocks with row permutations $\textbf{A'} = P A$. The permuted matrix is stored and passed to the execution phase, where our custom
CUDA kernel performs the matrix multiplication.

\subsection{Data Structures}
\label{sec:bcsr}
Our implementation leverages the widely used blocked format for sparse matrices, BCSR. In \Cref{fig:overview},
we provide an example of the BCSR format, which consists
of three arrays. Similarly to CSR, the BCSR array $rowPtr$ stores the offset pointers in the $col$ array for each block row,
and $col$ holds the column index for each block.
Since the blocks are stored as dense, every $h \cdot w$ consecutive values in array $val$ represent one block,
which may require filling some values with zeros (Figure~\ref{fig:overview}, top left).
Matrices stored in this format can directly be used as input for the MMA Units since the block dimensions $h$ and $w$ of BCSR 
match the dimensions of the MMA API calls. This, in turn, depends on the used precision: e.g., for FP16, we use the block
size $16 \times 8$, corresponding to the \textsc{m16n8k16} instruction (Listing~\ref{code:mma}).

It is worth noting that both DASP and cuSPARSE use the CSR format, while Magicube uses a Strided Row-major BCRS (SR-BCRS) format. The dense vectors in SR-BCRS are stored in a stride-wise row-major manner. If the number of dense vectors in the row is not a multiple-of-stride, zero vectors are padded for the last stride.


\subsection{Preprocessing}
\label{sec:reordering}

While finding the optimal block-minimizing permutation
is NP-hard, various heuristics exist.
We tested various state-of-the-art reordering schemes that cluster the non-zero values together to minimize the number of blocks. Reverse Cuthill–McKee~\cite{doi:10.1137/0713020} minimizes the matrix bandwidth. Saad's algorithm~\cite{saad2001} uses a similarity metric for clustering 
similar rows to increase the spatial locality. Çatalyürek et al.~\cite{10.1145/3571808} uses hypergraph partitioning techniques to minimize the cut between the rows. Zhao et al.~\cite{Zhao2020} uses
the Gray code ordering to maximize data locality. Sylos Labini et al.~\cite{10027183} use Jaccard's similarity metric to determine the blocking. In our tests, Sylos Labini's algorithm provided the best 
reduction in the block count. Thus, we use this as our baseline preprocessing routine. 

Sylos Labini's algorithm reduces the padding with zeros by clustering similar rows together. Intuitively, similar rows have many zero and nonzero values in the same columns. In order to measure the similarity of two rows $v, w$ they use the Jaccard distance $J(v, w) = 1 - \frac{|v \cap w|}{|v \cup w|} $, which measures the ratio of padding to the total cluster area.
Their algorithm iteratively performs the following greedy procedure: create a new cluster $c$ and choose an unclustered row $v$; for all other unclustered rows $w$ check whether the Jaccard distance $dist(w, pc)$ is less than a threshold distance ($pc$ represents the union of rows in cluster $c$); merge all rows for which the distance is less than the threshold.
This procedure is repeated until all rows belong to some cluster.

While Sylos Labini's algorithm performs only row permutation to avoid reshuffling the right-hand-size matrix $B$, we tested the impact of both row and column reordering to maximize the block reduction ratio.
We note that while row permutation comes with a minimal cost (the entire computation schedule is similar up to the permutation of the result matrix), permuting columns causes the overhead of accessing $B$, so any reduction in the number of blocks must be large enough to compensate for this. Our experiments (Section~\ref{sec:res_preprocessing}) show that performing additional column permutation 
does not provide sufficient benefits. Therefore, in our implementation, \modelname{} performs row-only permutation as a block-densification preprocessing step.

It is worth noting, that in some special cases, such as band matrices, the sparse matrix is already structured to minimize the number of non-zero blocks. In this case, the permutation $P$ becomes the identity matrix, and the sparsity pattern of the permuted matrix $\textbf{A'}$ is the same as for the input matrix $A$. 


\subsection{Implementation}
\label{sec:implementation}
\begin{algorithm}
\caption{Pseudocode of the warp-level \modelname{} kernel }\label{alg:pseudocode}
\begin{algorithmic}[1]
\Require $valuesBcsr, rowPtrBcsr, colIdxBcsR, B, C$
\For{$laneid = 0$ \textbf{to} $31$ \textbf{in parallel}}
    \State $RC \gets 0$
    \For{$blocks$ \textbf{in } $bcsrVals[row]$}
        \State $RD \gets RC$
        \State \textsc{memcpy\_async}($A\_shared$, $bcsrVals[idxA]$)
        \State \textsc{memcpy\_async}(B\_shared, $B[idxB]$)
        \State \textsc{ldmatrix\_x4}($RA$, $A\_shared$)
        \State \textsc{ldmatrix\_x2}($RB$, $B\_shared$)
        \State \textsc{hmma16816}($RD$, $RA$, $RB$, $RC$)
    \EndFor
    \State $C\_shared \gets RC$
    \State $C[idxC + laneid] \gets C\_shared$
\EndFor
\end{algorithmic}
\end{algorithm}

\lstset{language=C++, float, numbers=left, frame=tb, captionpos=b, basicstyle=\tt\scriptsize, numberstyle=\ssmall, numbersep=1pt, 
showstringspaces=false}

\begin{lstlisting}[float=h, caption={mma.m16n8k16 Tensor Core instruction in FP16}, label={code:mma}]
#define HMMA16816(RD0, RD1, RA0, RA1, RA2, RA3,\
    RB0, RB1, RC0, RC1) \
    asm volatile("mma.sync.aligned.m16n8k16.row.col.\
    f16.f16.f16.f16 \
    {%0, %1}, {%2, %3, %4, %5}, {%6, %7}, {%8, %9};\n" \
    : "=r"(RD0), "=r"(RD1)  \
    : "r"(RA0), "r"(RA1), "r"(RA2), "r"(RA3), \
      "r"(RB0), "r"(RB1), "r"(RC0), "r"(RC1))

\end{lstlisting}

\begin{lstlisting}[float=h, caption={LDMATRIX\_X2}, label={code:ldmatrix2}]
#define LDMATRIX_X2(R0, R1, addr) \
    asm volatile("ldmatrix.sync.aligned.x2.m8n8.shared.b16 \
    {%0, %1}, [%2];\n" : "=r"(R0), "=r"(R1) : "r"(addr))

\end{lstlisting} 

\begin{lstlisting}[float=h, caption={LDMATRIX\_X4}, label={code:ldmatrix4}]
#define LDMATRIX_X4(R0, R1, R2, R3, addr) \
    asm volatile("ldmatrix.sync.aligned.x4.m8n8.shared.b16 \
    {%0, %1, %2, %3}, [%4];\n" \
    : "=r"(R0), "=r"(R1), "=r"(R2), "=r"(R3) \
    : "r"(addr))

\end{lstlisting}

We now describe selected most important optimization steps employed in our library. These steps are illustrated in Figure~\ref{fig:perf_model}. Full code is publicly available on GitHub\footnote{\url{https://github.com/spcl/smat}}.

\noindent
\textbf{T: TC API} We execute in half precision $mma.m16n8k16$ operation. Listing \ref{code:mma} shows the PTX code example we incorporate in our implementation for using TCs. 

\noindent
\textbf{B: BCSR iteration} We use arrays \texttt{rowPtr} and \texttt{colIdx} from \Cref{fig:overview} in order to iterate only over non-zero blocks. Without these data structures, one needs to iterate over every block and check whether that block is non-zero. More details can be found in \Cref{alg:pseudocode} and \Cref{sec:bcsr}.

\noindent
\textbf{C: collective loads} \texttt{cuda::memcpy\_async} achieves overlapping computation with data movement, allowing for faster transfer of data from global memory to registers.. \Cref{sec:memcpy_async} contains additional information.

\noindent
\Cref{alg:pseudocode} presents the pseudocode for the CUDA kernel of \modelname{}. Each warp is responsible for the calculation of a submatrix of matrix $\textbf{C}$ such that the dimensions correspond to the dimensions of the TC. As the dense matrix $\textbf{B}$ has dimension $N << K$, most of its width of $\textbf{B}$ is loaded in the memory. Afterwards, non-zero blocks using the BCSR format are loaded from the global memory into registers in an efficient way. For that, we use \texttt{cuda:memcpy:async} for loading into shared memory, and the PTX command \texttt{ldmatrix} shown in Listing 2 and 3 for loading into registers in the required format. Then the execution of the TCs is called with the command \texttt{HMMA16816}. Finally, the result stored in the registers is transferred back into global memory.


\subsection{Asynchronous Data Loads}
\label{sec:memcpy_async}
In order to hide the latency of transferring data from the global GPU memory to 
shared memory we utilize \texttt{cuda::memcpy\_async}.
These asynchronicity features enable overlapping computations with data movement, reducing total execution time. The \texttt{cudaMemcpyAsync} function allows data movement between CPU memory and GPU global memory to be overlapped with kernel execution. Similarly, the \texttt{cuda::memcpy\_async} function allows data movement from GPU global memory to shared memory to be overlapped with thread execution.
It is important to note that copying data from global to shared memory without \texttt{cuda::memcpy\_async} is a two-step process.   The process of copying data from global memory into registers and then from registers into shared memory is performed in multiple
stages through the memory hierarchy. To avoid this, \texttt{cuda::memcpy\_async} can be used to directly transfer data from global memory to shared memory using DMA engines without involving registers. This frees up the thread block from the task of moving data and allows registers to be used for computations.



\section{Evaluation}
\label{sec:evaluation}

We note that, to the best of our knowledge, there is no high-performance library for unstructured SpMM using tensor cores that 
would simultaneously work on real-world, highly sparse ($>$90\%), and irregular matrices while utilizing Tensor Cores efficiently. Existing solutions tend to either focus on the former (highly sparse unstructured matrices in the CSR format~\cite{naumov2010cusparse}) or the latter (utilizing TC for relatively dense, small, structured matrices in very low precision commonly found in Machine Learning (ML)~\cite{Magicube, lu2023dasp, castro2023venom, kim2022analysis}).
We compare \modelname's performance against existing solutions that can be employed in such a scenario, exposing their 
weaknesses on a large set of both real-world and synthetic matrices of varying size, structure, and sparsity.
Furthermore, we showcase the effectiveness of the preprocessing permutation of sparse matrices. 

We also test different solutions for our preprocessing step to minimize the number of blocks. Our experiments show, that Sylos Labini's algorithm
(Section~\ref{sec:reordering}) performs the best for our test matrices. We evaluate two variants: the original version proposed by the authors
that permutes only the rows, and our experiment on both row and column permutation.

\subsection{Comparison Targets}
\label{sec:targets}
As comparison targets, we use Magicube, which is a high-performance sparse-matrix library for low-precision integers on Tensor cores
optimized for ML scenarios.
For a fair comparison, we evaluate the Magicube\footnote{\url{https://github.com/Shigangli/Magicube}} mixed precision int16 since the 
throughput is equal to fp16 on TC.
Although DASP\footnote{\url{https://github.com/SuperScientificSoftwareLaboratory/DASP}} focuses on SpMV, we consider it as a batched vector algorithm by iteratively performing SpMV --- for this reason, we focus mostly on tall-and-skinny dense matrices.
Finally, we compare against a cuSPARSE implementation of SpMM using the CSR format\footnote{\url{https://github.com/NVIDIA/CUDALibrarySamples/blob/master/cuSPARSE}}.

\subsection{Hardware Infrastructure}
We run our experiments on the Swiss National Computing Center's Ault compute cluster. Each node is equipped with a single NVIDIA A100-SXM4-40GB GPU, and AMD EPYC 7742 @ 2.25GHz CPU. The A100 driver version is 530.30.02.

\subsection{Software Stack}
All experiments were executed using the GCC 12.3.0 compiler, NVIDIA nvcc v12.0, NVIDIA cuSPARSE v12.0, NVIDIA CUDA Toolkit v12.0, Python 3.9, and the following Python libraries: Pandas, Matplotlib, Numpy, Scipy, and Seaborn.

\begin{table}
\centering
\begin{tabular}{lllll}
\hline
\textbf{Domain} &\textbf{Name}      & \textbf{Size} & \textbf{\emph{nnz}}  &  \textbf{Sparsity}\\ \hline
optimization & mip1               & 66K$\times$66K            & 10.4M         &  99.76\%\\  
quantum chem. & conf5\_4-8x8    & 49K$\times$49K            & 1.9M      &  99.92\%    \\ 
2D/3D mesh & cant               & 62K$\times$62K            & 4M         &  99.89\%   \\ 
weighted graph & pdb1HYS            & 36K$\times$36K            & 4.3M         &  99.67\% \\ 
fluid dynamics & rma10              & 46.8K$\times$46.8K        & 2.3M     &  99.89\%     \\ 
2D/3D mesh & cop20k\_A          & 121K$\times$121K          & 2.6M       &  99.98\%   \\ 
2D/3D mesh & consph             & 83K$\times$83K            & 6M          &  99.91\%  \\ 
structural & shipsec1           & 140K$\times$140K          & 7.8M          &  99.96\% \\ 
circuit simulation & dc2                & 116K$\times$116K          & 766K    &  99.99\%      \\ \hline
\end{tabular}
\vspace{0.3cm}
\caption{Selected matrices from the SuiteSparse}
\label{tab:matrix_sizes}
\end{table}

\subsection{Tested Scenarios}
\label{sec:scenarios}
As mentioned in Section~\ref{sec:targets}, neither DASP nor Magicube are optimized for large SpMM. Magicube, 
which is optimized for relatively small matrices found in ML, has a large memory footprint. This significantly
limits the number of matrices we can run all libraries on.
We consider two different origins of matrices:
\begin{itemize}
    \item \textbf{SuiteSparse Collection}~\cite{Kolodziej2019}: to make a fair comparison against DASP,
    we use the same representative set of matrices as used by Lu et al.~\cite{lu2023dasp}. These matrices come from 7 application domains, and are considered a broad representation of different types of sparse matrices.
    Out of 21 matrices used by DASP, Magicube can support only 9 of them (\Cref{tab:matrix_sizes}).
    \item \textbf{Synthetic band matrices}: we generate a series of band matrices of variable bandwidth, scaling their
    sparsity from 99.7\% all the way to dense matrices (0\% sparsity),
    These matrices serve as the empirical evidence for the claims made in \Cref{sec:performance_model}. \linebreak
    \textbf{\emph{Motivation:}} Testing general-purpose unstructured SpMV/SpMM routines on highly regular matrices
    is common among the HPC benchmarks. HPCG~\cite{luszczek2006hpc} -- one of the main benchmarks to rank supercomputers -- tests this scenario:
    high performance SpMV on a matrix originating from a 3D grid computation.
\end{itemize}

\subsection{Methodology}
We first execute a warm-up run for all the experiments to obtain reliable measurements of the execution time on the GPU. Afterward, for each method, we run the kernel call 10 times and report the arithmetic mean. 
We observe that the variance in time measurements is very low: across all experiments, the Coefficient of Variation $CV = \sigma / \mu = 0.0182$ 
(geometric mean across all the runs).
 For example, \modelname{} wallclock time on \texttt{cop20k\_A} is $0.125$ms, and the variance is $1$ns.
 Therefore, when plotting the measurements, we do \textbf{not include the confidence intervals} nor the standard deviation, as it would be 
 unreadably small and would only clutter the display.

\section{Results}
\label{sec:results_cusparse}

\begin{figure*}[h]
\centering
\centerline{\includegraphics[width=2\columnwidth]{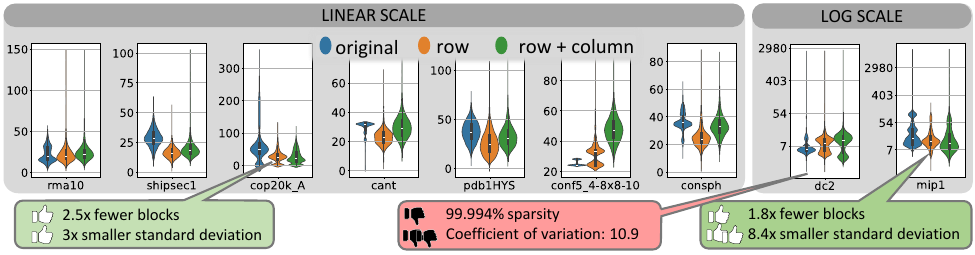}}
\caption{Distribution of the blocks count per row in the  BCSR format in the input matrix (original), after 
the row reordering, and after row and column reordering for the test matrices. For \texttt{cop20k\_A}, 
row reordering reduces the number of BCSR blocks by 2.5x and the standard deviation by 3x. For \texttt{mip1},
while the reduction of the total block count is slightly smaller (1.8x), the standard deviation is reduced
by 8.4x, significantly improving the load balance for our 2D parallel schedule. Matrix \texttt{dc2} is the most
adversarial for \modelname{}: with its extreme sparsity and power-law distribution of nonzeros per row,
the runtime cannot utilize tensor cores, and the warp-level static schedule generates high load imbalance on SMs.
}
\label{fig:violin}
\end{figure*}

\subsection{Preprocessing reordering}
\label{sec:res_preprocessing}

\begin{figure}[t]
\centering
\centerline{\includegraphics[scale=0.202]{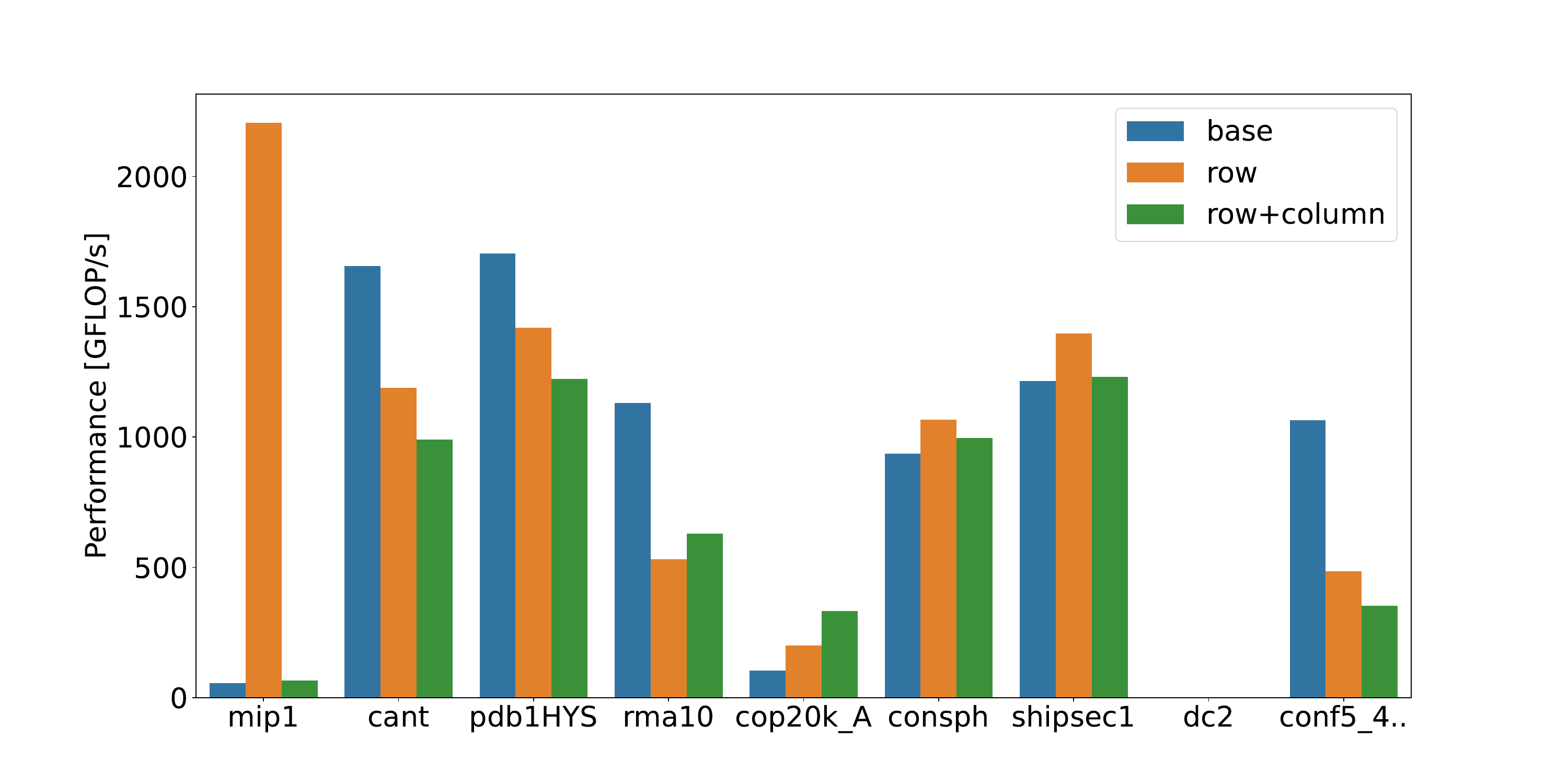}}
\caption{Reordering effect on the performance of \modelname{} on 9 representative matrices from SuiteSparse.}
\label{fig:ordering_smatel}
\end{figure}

\begin{figure}[t]
\centering
\centerline{\includegraphics[scale=0.202]{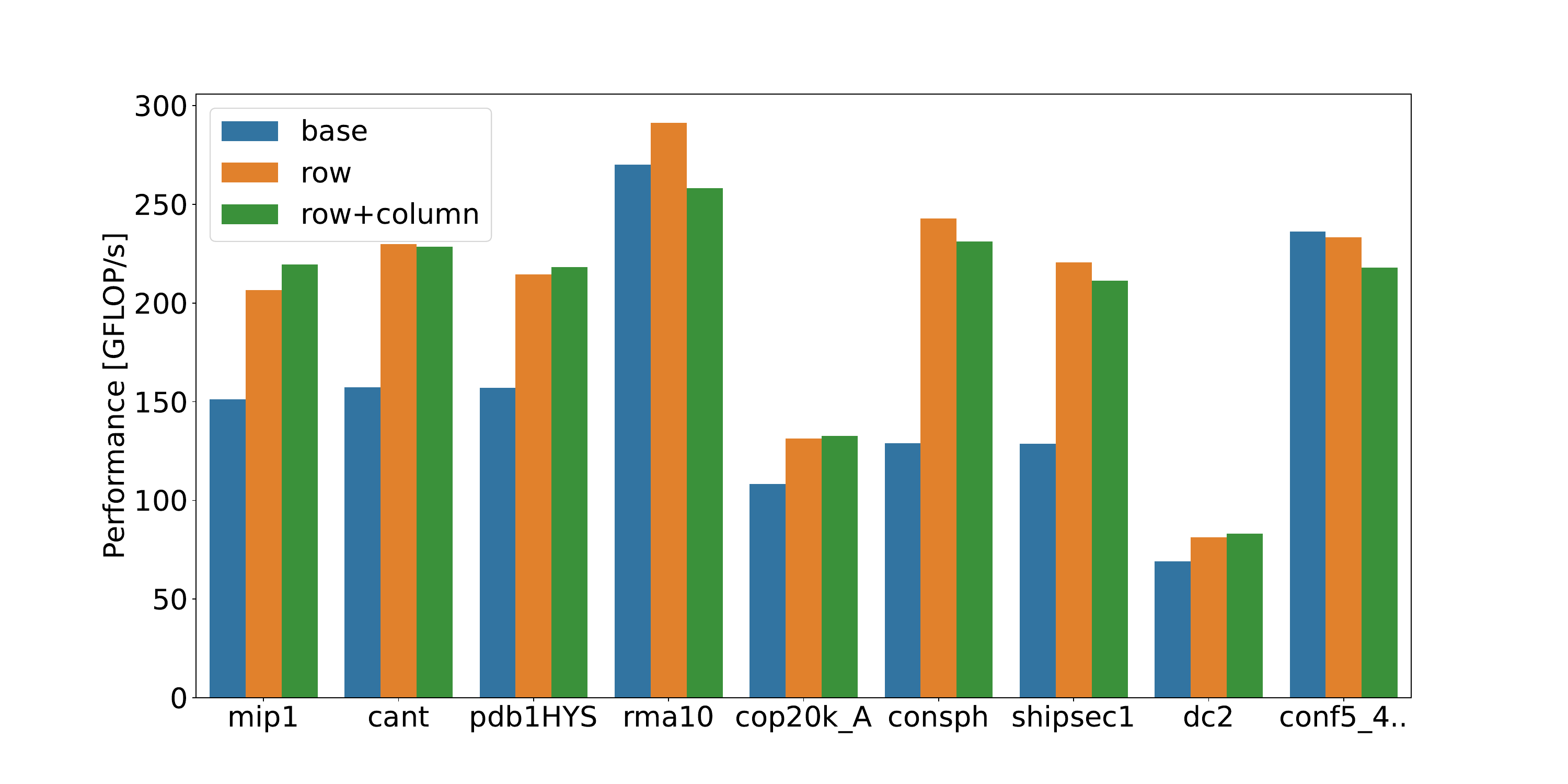}}
\caption{Reordering effect on the performance of DASP on 9 representative matrices from SuiteSparse.}
\label{fig:ordering_dasp}
\end{figure}

\begin{figure}[t]
\centering
\centerline{\includegraphics[scale=0.202]{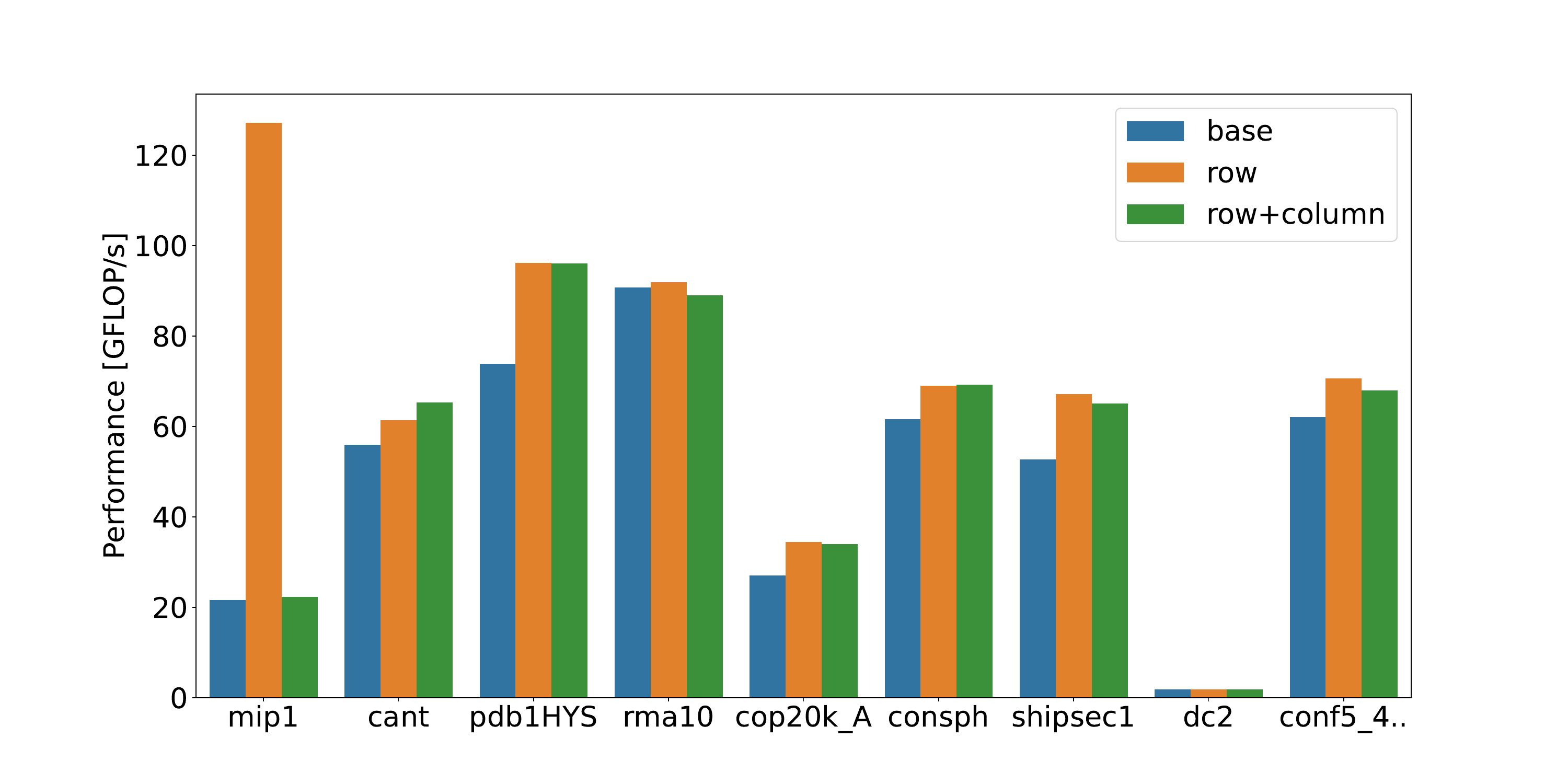}}
\caption{Reordering effect on the performance of Magicube on 9 representative matrices from SuiteSparse.}
\label{fig:ordering_magicube}
\end{figure}

\begin{figure}[t]
\centering
\centerline{\includegraphics[scale=0.202]{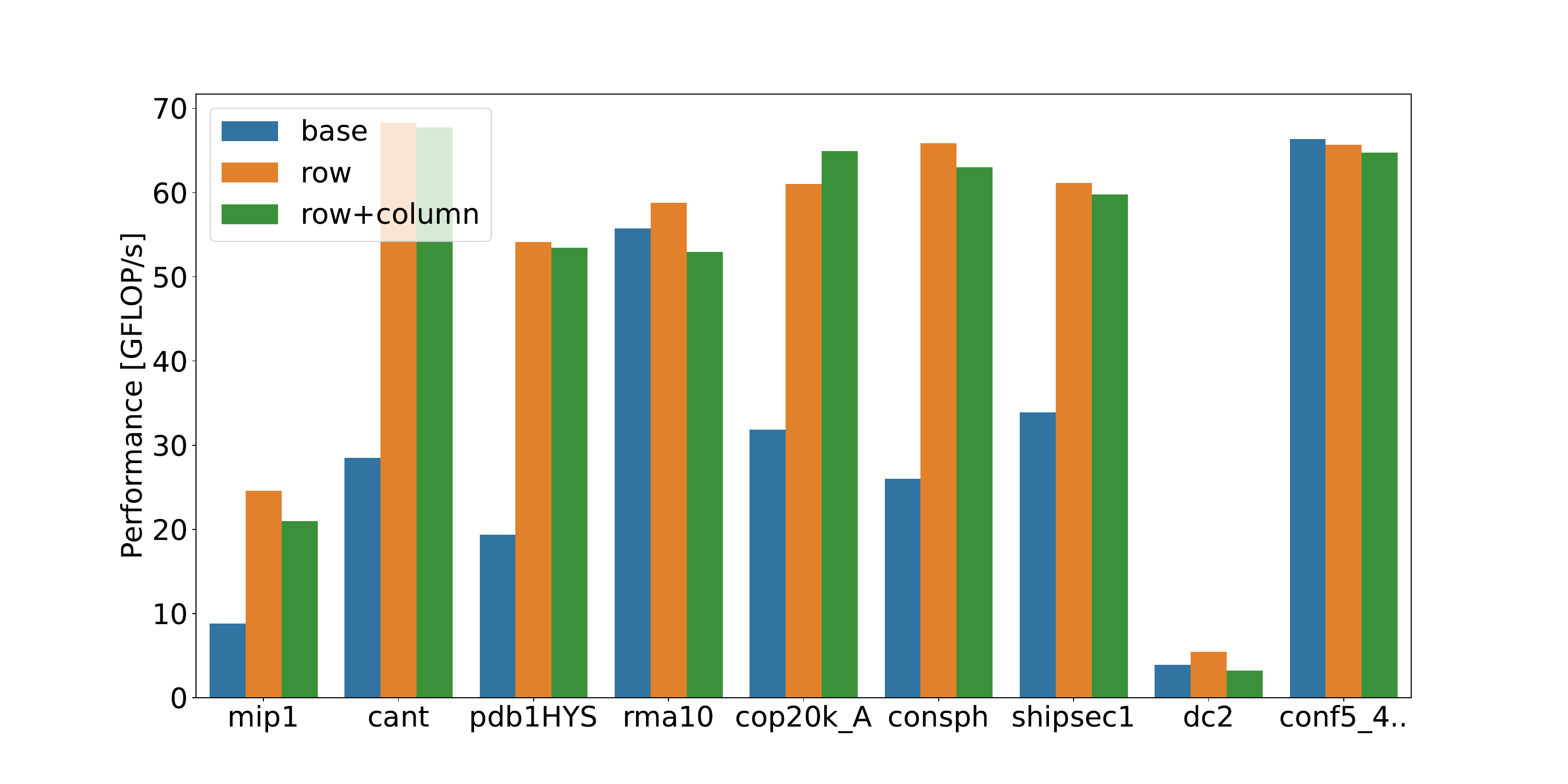}}
\caption{Reordering effect on the performance of cuSPARSE on 9 representative matrices from SuiteSparse.}
\label{fig:ordering_cusparse}
\end{figure}

We start by analyzing the impact of reordering matrices on computation
time for representative matrices from the SuiteSparse Collection. 
 Figure~\ref{fig:ordering_smatel} demonstrates the importance of reordering for \modelname{}.
 As the performance positively correlates to the number of blocks in the reordering,
 this confirms the performance model introduced in \Cref{sec:performance_model}.
 We observe that some input matrices are already well-structured: e.g., \linebreak \texttt{conf5\_4-8x8}, which originates from the 
 quantum chemistry simulations, is a sparse band matrix -- the locality of interactions between particles put all nonzeros
 relatively close to the diagonal. In this case, the Jaccard's similarity metric used by our preprocessor 
 incorrectly reshuffles the rows, \emph{increasing} the number of blocks.
We evaluate the same setting for DASP (Figure~\ref{fig:ordering_dasp}), cuSPARSE (Figure~\ref{fig:ordering_cusparse}),
and Magicube (Figure~\ref{fig:ordering_magicube}).
In general, the reordering decreased the number of BCSR blocks in 6 our of 9 matrices, with the block count reduction
ranging from $1.3\times$ (\texttt{cant}) up to $2.4\times$ (\texttt{cop20k\_A}). This significantly translates
to the final performance solution, as discussed in Section~\ref{sec:res_suitesparse}, confirming 
the importance of this preprocessing step.
The distribution of the number of 
blocks per row in BCSR for the input matrices, after the row permutation, 
and after row and column permutation for all tested matrices can be found in Figure~\ref{fig:violin}.

\subsection{SuiteSparse}
\label{sec:res_suitesparse}

\begin{figure*}[htbp]
\centering
\centerline{\includegraphics[width=2\columnwidth]{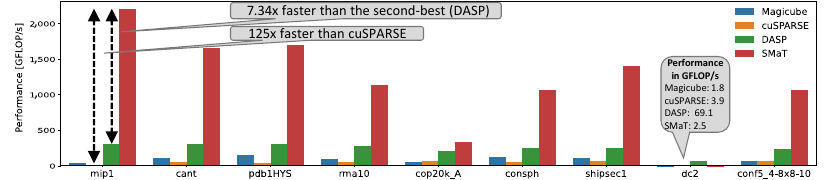}}
\caption{Performance comparison on 9 representative matrices from SuiteSparse.}
\label{fig:suitesparse_performance}
\end{figure*}

As stated in Section~\ref{sec:scenarios}, we measure the performance in GFLOP/s on 9 matrices
with the non-zero count ranging from $766k$ to $10.4M$ and the number
or rows ranging from $M=36k$ to $M=140k$. While \modelname{} supports much bigger matrices as well,
Magicube is designed to optimize ML workloads, which host much smaller matrices. In our experiments
Magicube's internal preprocessing and representation runs out of memory for larger matrices from 
SuiteSparse. Therefore, we include only matrices that are supported by all comparison targets. 
Furthermore, since we are using batched SpMV to simulate SpMM for DASP, we keep the batch dimension
small ($N=8$): large values of $N$ would make 
a batched-SpMV less competitive compared to the direct SpMM routines, as discussed in \Cref{sec:matrix_dimension}.

The dimensions and number of non-zero elements of the benchmark matrices are shown in \Cref{tab:matrix_sizes}. 
Across 9 matrices \modelname{} is better on average $7.71\times$ (geometric mean) than the respective baselines. Compared to baselines \modelname{} is: $2.60\times$ faster (up to $7.34\times$) than DASP, $10.78\times$ faster (up to $51.23\times$) than Magciube, $16.32\times$ faster (up to $125.48\times$) than cuSPARSE. The results are presented in \Cref{fig:suitesparse_performance}.

\noindent
\textbf{Best case scenario.}
\modelname{} has the largest performance improvement compared to baselines on the \texttt{mip1} matrix.
This use-case emphasizes the importance of good preprocessing that simultaneously minimizes 
the number of blocks \emph{and} the row load-imbalance. As discussed in Section~\ref{sec:reordering},
our preprocessing routine is especially effective for \texttt{mip1}, reducing the total
number of blocks by 1.8 times, but more importantly, the standard deviation of block count per row
from 146.2 to 17.4. This improves the load balance by 8.4 times (Figure~\ref{fig:violin}),
which is crucial for an efficient parallel schedule. Interestingly, other libraries
seem not to be able to take full advantage of this property (Figures~\ref{fig:ordering_dasp},~\ref{fig:ordering_magicube},~\ref{fig:ordering_cusparse}). Since they also employ
their internal preprocessing algorithms, we expect that in this use-case their
preprocessor actually \emph{increases} the load imbalance. However, a detailed analysis
of this aspect is out of scope of this work.

\noindent
\textbf{Worst case scenario.} 
We observe that \modelname{} does not always perform the best. \texttt{dc2} is the sparsest among 
tested matrices (sparsity 99.994\%) with a very high row imbalance: the standard deviation 
of the number of blocks per row is 169.9, while the arithmetic mean is
15.65. This makes \texttt{dc2} especially ill-suited for \modelname{}'s execution model with 
static 2D parallel schedule, with most of the blocks containing only a single non-zero value,
heavily underutilizing tensor cores, achieving only 2.5 GFLOP/s. 
This is the use-case where either DASP row-packing algorithm
benefits most (69.1 GFLOP/s) or even a non-blocked CSR used in cuSPARSE (3.9 GFLOP/s).

\subsection{Synthetic Matrices}
\label{sec:results_synthetic}
Formally, an $n\times n$ matrix $\textbf{A} = (a_{i,j})$ is a band matrix if all elements are zero outside a diagonally bordered band of width $b$: $a_{i,j} = 0 \quad \text{if} \quad j < i - b \quad \text{or} \quad j >  i + b; \quad b \geq 0.$
We evaluate performance on band matrices to measure the dependence of \modelname{} on the number of blocks $n_e$ while isolating the randomness of the sparse matrix structure. 
Furthermore, for band matrices, blocks in BCSR format are already dense, so no further reordering is necessary.
Hence, we do not incorporate the effect of the reordering while evaluating the kernel. 

We perform our tests on matrix $A$ of size $16,384\times 16,384$ with varying the bandwidth from $b=64$ all the way up to $b=16,384$, 
effectively making the matrix dense. This allows us to measure a very important performance factor in sparse computations: 
\textbf{at what sparsity threshold a sparse library can outperform a highly-optimized dense library, if the matrix is explicitly
padded with zeros?} Previous experiments~\cite{sparsert} suggest that this can be as high as 99\%. 
While a band matrix is definitely \emph{not} a fair representation of an unstructured sparse matrix, we show that in this
artificial scenario, \modelname{} can be competitive with cuBLAS. 

\noindent
\textbf{Note:} We measure the performance of cuBLAS only once - for a dense matrix $16,384 \times 16,384$. 
We then report cuBLAS performance as the \emph{effective} FLOP/s, that is, we scale it by the fraction of nonzeros.

\begin{figure*}
    \centering
    \subfloat[
    $N=8$
    ]
    {
    \hspace*{-0.4cm}\includegraphics[width=1.05\columnwidth]{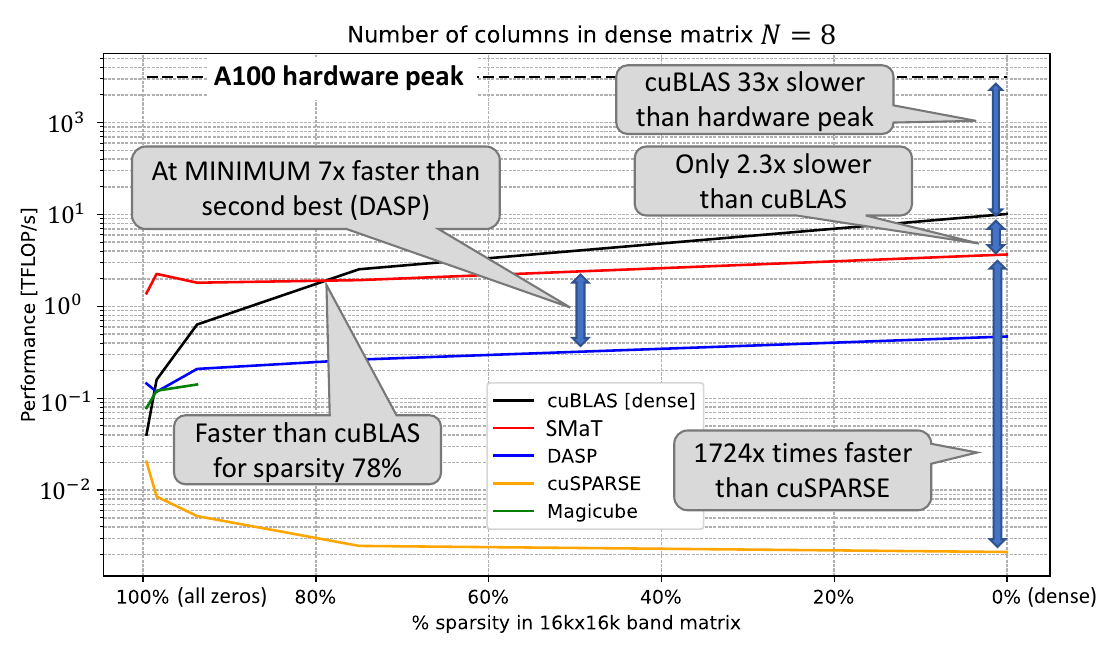}
    }
    \subfloat[
    $N=128$
    ]
    {
    \includegraphics[width=1.1\columnwidth]{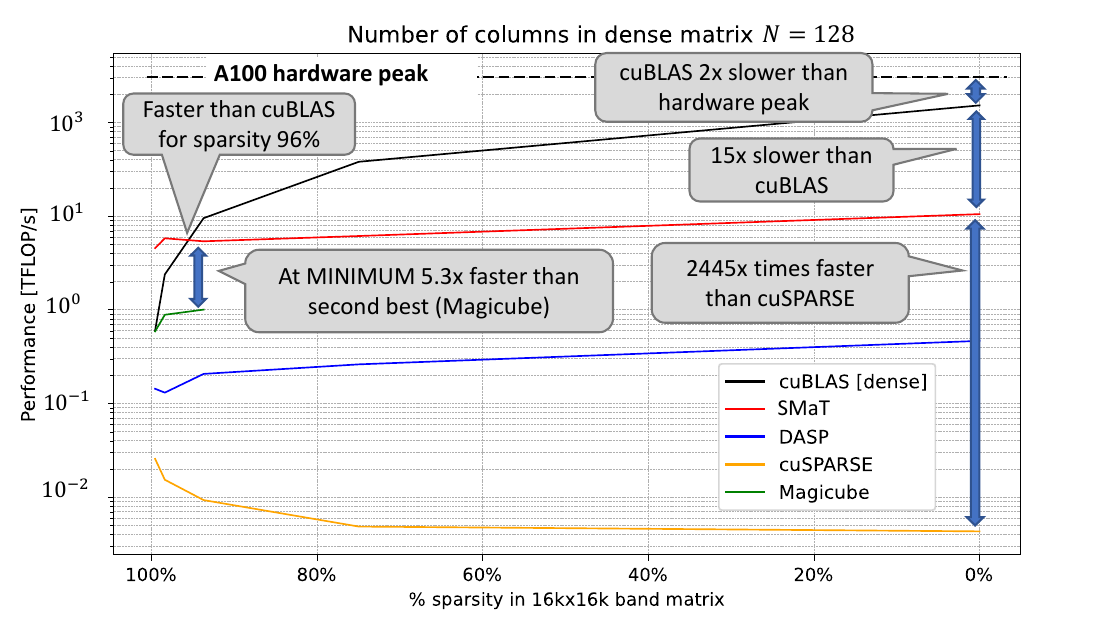}
    }
    \caption{ Measured performance of multiplying a synthetic band matrix $16k \times 16k$ with dense matrix $16k \times N$. 
    Bandwidth $b$ varies from $b=64$ to $b=16k$. Corresponding sparsity ranges from 99.7\% to  0\% (fully dense matrix). }
    \label{fig:synthetic}
\end{figure*}

We first measure the performance of $C=AB$ with the number of columns $N=8$ and present the results in \Cref{fig:synthetic}a.
\modelname{} is up to $1,724\times$ faster than cuSPARSE and is at minimum $7\times$ faster than the second best method DASP. We notice that \modelname{} outperforms cuBLAS for sparsity $\geq 78\%$ and is only $2.3\times$ slower than cuBLAS in the dense case.

For $N=128$, the difference between \modelname{} and the baselines rises (\Cref{fig:synthetic}b).
As $N$ grows, cuBLAS performance increases and reaches closer to the hardware peak in the dense setting. Nevertheless, \modelname{} performs better than cuBLAS for sparsity $\geq96\%$. Compared with other methods, \modelname{} is at minimum $5.3\times$ faster than the second best method, Magicube, and gets up to $2,445\times$ faster than cuSPARSE.

\subsection{Scaling Matrix Dimensions}

\label{sec:matrix_dimension}
\begin{figure} 
\centerline{\includegraphics[width=\columnwidth]{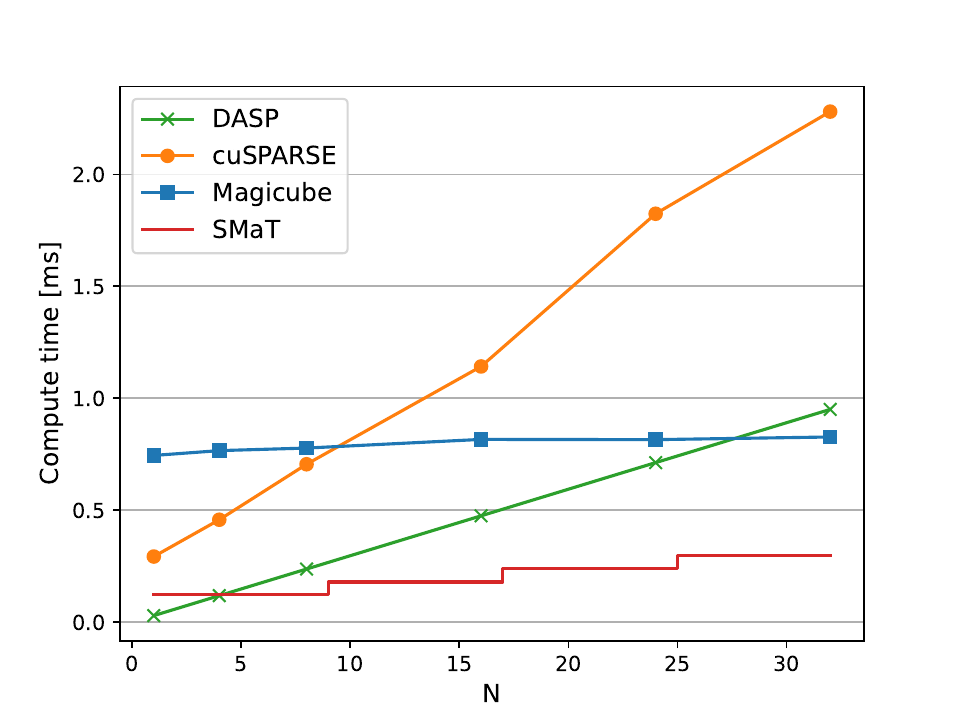}}
\caption{Wall-clock time of SpMM $AB=C$, where $A$ is the sparse matrix \textit{cop20k\_A} and $B$ is a tall-and-skinny dense matrix 
with varying number of columns $N$.}
\label{fig:matrix_dimension}
\end{figure}

\Cref{fig:matrix_dimension} demonstrates the relationship between the compute time and the outer dimension $N$ of the dense matrix $\textbf{B}$. 
The performance evaluation is conducted on the sparse matrix \textit{cop20K\_A} from \Cref{tab:matrix_sizes}. 

As $N$ grows, \modelname{} exhibits the best performance among
all the baselines. While DASP and cuSPARSE exhibit a degradation in performance as $N$ increases, Magicube, similar to \modelname{}, shows a
slow increase. Although DASP remains the fastest for $N =
1$, i.e., for SpMV.

For larger dimensions, such as $N = 1,000$, \modelname{} outperforms the baseline methods in terms of execution time.
Specifically, \modelname{} takes $6.98$ms to execute, while Magicube takes
$12.13$ms, DASP takes $29.70$ms, and cuSPARSE takes $60.07$ms. This means that
\modelname{} is $1.73\times$, $4.24\times$, and $8.60\times$ faster than the baseline
methods, respectively.



\subsection{Distribution of the Number of Blocks per Row} 
The reordering experiments reveal an important property 
of our 2D parallel BCSR schedule. By using a fixed parallel grid and assigning one block of the output matrix per warp
(Figure~\ref{fig:overview}), \modelname{}'s schedule is sensitive to the highly skewed distribution of blocks per row --- some 
warps can have significantly more non-zero blocks to process than the others. This can be viewed, for example, for matrix 
\texttt{cant}: while the reordering decreases the mean number of blocks per row from 29.6 to 22.8, 
it \emph{increases} its standard deviation from 4 to 5.2 (Figure~\ref{fig:violin}). This results in the actual decrease
in \modelname{}'s performance (Figure~\ref{fig:ordering_smatel}). In contrast, the remaining libraries benefit from the 
reduced block count despite the increased load imbalance (Figures~\ref{fig:ordering_dasp},~\ref{fig:ordering_magicube}, and~\ref{fig:ordering_cusparse}).

\subsection{Results Summary}
Our results consistently show significant improvement over existing solutions. Several observations stand out from our experiments:
\begin{itemize}
    \item DASP, which is a highly optimized SpMV library, often performs a single SpMV \emph{slower} than \modelname{} performs a tall-and-skinny SpMM with $N=8$,
    \item for relatively dense matrices, cuSPARSE performance is low (up to 2,445 times lower than \modelname{}). We also 
    observe that its performance \emph{drops} for denser matrices (Figure~\ref{fig:synthetic}),
    \item Magicube is a highly-specialized library for ML: while being the second-fastest in the band matrix experiment (Figure~\ref{fig:synthetic}b), it quickly runs out of memory
    for larger values of $N$, making it unsuitable for general-purpose SpMM kernels,
    \item Sparse matrix column permutation preprocessing does not significantly reduce the number of blocks in BCSR format.
\end{itemize}

\section{Related Work}
\newparagraph{\textbf{GEMM}} Dense matrix multiplication is an active field of research for decades. Cannon's algorithm~\cite{cannon1969cellular} was the first distributed 2D parallel algorithm for square matrices. 
van de Geijn et al. extended it to non-square matrices in the SUMMA algorithm~\cite{van1997summa}. 
Agarwal et al. presented the distributed 3D algorithm that parallelizes also the reduction dimension~\cite{agarwal1995three}.
Kwasniewski et al. introduced COSMA~\cite{cosma} -- a communication optimal 2.5D parallel GEMM library for both CPUs and GPUs.
Research in this area varies depending on the hardware platforms for which it is designed (GPU or CPU), the type of sparsity it addresses (structured or unstructured), and the precision levels it targets, including formats such as fp16, fp32, fp64, int8, and other precision
schemes.

\newparagraph{\textbf{Tensor Cores}}
In addition to GEMM and its applications such as machine learning~\cite{castro2023venom}, Tensor cores
are successfully employed to enhance basic operators such as scan and reduction~\cite{reductionScan}, 
stencil computation~\cite{stencil}, and FFT~\cite{fft}.

\newparagraph{\textbf{Reordering}}
Reordering and blocking have been explored to accelerate parallel multiplication, particularly in SpMV~\cite{reorderingSpmv} and SpMM~\cite{reorderingSpmm}.  Row-reordering is often used to promote efficient memory hierarchy usage by enhancing locality~\cite{paoloReordering}. This involves moving rows with similar nonzero structures closer together to locally densify blocks and reduce cache misses during SpMV or SpMM~\cite{cacheMisses}.

\newparagraph{\textbf{SpMV}}
One line of research focuses on multiplying sparse matrices with dense vectors (SpMV). 
Researchers have explored the trade-offs between balancing workload~\cite{balancingWorkload, balancingWorkload2}, data locality~\cite{dataLocality}, and format generation~\cite{formatGen1,du2022alphasparse}. SpMV overview studies can be found in~\cite{goumas2009performance}.

\newparagraph{\textbf{SpMM}}
Gao et al. presents an overview of the existing research on sparse matrix multiplication (SpGEMM)~\cite{overviewGEMM}.
NVIDIA has developed cuSPARSE~\cite{naumov2010cusparse} and cuSPARSELt~\cite{cusparselt} for sparse matrix multiplication. cuSPARSE
is capable of performing unstructured SpMM for sparsity above 95\% on CUDA Cores, while cuSPARSELt
uses Tensor
Cores and utilizes 2:4 structured sparsity.
Gale et al. design Sputnik~\cite{sputnik}, a GPU kernel to accelerate sparse matrix operations in neural networks. Chen et al. present vectorSparse~\cite{vectorSparse}, proposing column-vector-sparse-encoding for SpMM. Li et al. develop Magicube~\cite{Magicube}, a high-performance library for low-precision integers on Tensor Cores focusing on deep learning as well. Furthermore, Li et al. propose the SR-BCRS format, based on the CS format, which is suitable for Tensor Cores. While Magicube is for structured sparse formats, Xue et al~\cite{fullPotential}. introduce a method for unstructured sparsity in fp16 precision.

\section{Conclusions}

This paper introduces \modelname{}, a novel general-purpose SpMM library that utilizes 
Tensor Core hardware units to accelerate Sparse Matrix-Matrix Multiplication (SpMM).
\modelname{} first employs modern state-of-the-art preprocessing techniques 
to significantly reduce the number of blocks in BCSR format, as well as improve the load balance
for the bottom-up 2D parallel schedule. The optimized CUDA implementation uses low-level API 
to fully utilize Tensor Cores, asynchronous shared memory loads, and warp-level memory alignment. 
We develop the empirical performance model that quantifies the benefit of both the preprocessing
and the kernel optimizations.

We perform a comprehensive performance study on both real-world sparse matrices from SuiteSparse
as well as synthetic matrices. Evaluated on NVIDIA Ampere GPUs, we show the supremacy of \modelname{}
compared to state-of-the-art libraries: Magicube, DASP, and cuSPARSE. We measure up to 7.3x speedup
vs the second-best library (2.6x on average), with up to 125x speedup over vendor-optimized cuSPARSE.

Furthermore, while scaling the sparsity ratio on the synthetic matrices for tall-and-skinny SpMM, 
we outperform cuBLAS for sparsity as low as 78\%, outperforming cuSPARSE by up to 2,445 times.
The presented results confirm that SpMM libraries do not need to be highly specialized to narrow 
use cases for fixed data types, sparsity patterns, and matrix sizes. 



\section*{Acknowledgments}
This research is carried out in the frame of the “UrbanTwin: An urban digital twin for climate action: Assessing policies and solutions for energy, water and infrastructure” project with the financial support of the ETH-Domain Joint Initiative program in the Strategic Area Energy, Climate and Sustainable Environment.
This work received EuroHPC-JU funding with support from
the European Union’s Horizon 2020 program and the
European Research Council under grant agreement PSAP,
number 101002047. We also wish to acknowledge the support
from the DEEP-SEA project under grant agreement number 955606.
The authors would like to thank the Swiss National Supercomputing Centre
(CSCS) for access and support of the computational resources.

\bibliographystyle{IEEEtran}
\bibliography{TC-SpMM.bib}

\end{document}